\documentclass{article}                  
\usepackage{opex2}

\usepackage{epsfig}
\begin{document}

\newcommand{\be}{\begin{equation}}
\newcommand{\ee}{\end{equation}}
\newcommand{\bc}{\begin{center}}
\newcommand{\ec}{\end{center}}
\newcommand{\bea}{\begin{eqnarray}}
\newcommand{\eea}{\end{eqnarray}}
\newcommand{\da}{\dagger}
\newcommand{\dg}[1]{\mbox{${#1}^{\dagger}$}}
\newcommand{\hlf}{\mbox{$1\over2$}}
\newcommand{\lfrac}[2]{\mbox{${#1}\over{#2}$}}
\newcommand{\nsz}[1]{\mbox{\normalsize ${#1}$}}
\newcommand{\lgsz}[1]{\mbox{\Huge ${#1}$}}
\newcommand{\ep}{\epsilon(t)}
\newcommand{\epo}{\epsilon_o}
\newcommand{\eps}{\epsilon^*(t)}
\newcommand{\epso}{\epsilon_o^*}
\newcommand{\epd}{\dot{\epsilon}(t)}
\newcommand{\epdo}{\dot{\epsilon}_o}
\newcommand{\epsd}{\dot{\epsilon}^*(t)}
\newcommand{\epsdo}{\dot{\epsilon}_o^*}
\newcommand{\QR}{\langle Q \rangle}
\newcommand{\QRT}{\langle Q^2 \rangle}
\newcommand{\PR}{\langle P \rangle}
\newcommand{\PRT}{\langle P^2 \rangle}
\newcommand{\x}{{\textsf x}}
\newcommand{\y}{{\textsf y}}
\newcommand{\z}{{\textsf z}}
\newcommand{\tn}{{\textsf t}}
\newcommand{\px}{\partial_{\textsf x}}
\newcommand{\py}{\partial_{\textsf y}}
\newcommand{\pz}{\partial_{\textsf z}}

\title{Symmetries and solutions 
of the three-dimensional Paul trap }

\author{Michael Martin Nieto}
\address{Theoretical Division (MS-B285), Los Alamos National Laboratory, 
University of California,
Los Alamos, New Mexico 87545, U.S.A.}
\email{mmn@lanl.gov}

\smallskip

\author{D. Rodney Truax}
\address{Department of Chemistry, 
 University of Calgary, 
Calgary, Alberta T2N 1N4, Canada}
\email{ truax@ucalgary.ca}

\begin{abstract}
Using the symmetries of the three-dimensional Paul trap, 
we derive the solutions of the 
time-dependent Schr\"odinger equation for this system, in both
Cartesian and cylindrical coordinates.  Our symmetry 
calculations provide insights that are not always obvious
from the conventional viewpoint. 
\end{abstract}




\section{Introduction}

I first got to know of Joe (the speaker was MMN)
because of his paper with Singh on the
time-energy uncertainty relation \cite{esingh}. [See the Appendix for this
story.]  To use the language of the discussions of this conference, 
the Eberly-Singh paper 
was ``fundamental'' though maybe not ``useful.''  But it certainly was
fun.  That is what we are presenting today.  
Something that is fun.  We are
going to discuss the Paul trap from a different point of view.  
We will discuss the Paul trap's symmetries and use them to obtain the
time-dependent solutions, without having to solve the mixed, second-order, 
partial, differential Schr\"odinger equation.   We will 
be exemplifying something that Christopher Gerry alluded to yesterday: 
although the Heisenberg and Schr\"odinger formulations of quantum
mechanics are equivalent, different things can be  more transparent in one
of them.  It was Pauli who taught us this.  
 

\section{The Paul trap}

The Paul trap provides a dynamically stable environment for charged particles
\cite{dawson}-\cite{paulnobel} and has been widely used in fields from  
from quantum optics to particle physics. 

Paul gives a delightful mechanical analogy \cite{paulnobel}.  Think 
of a mechanical ball put at the center of a saddle surface.  With no 
motion of the surface, it will fall off of the saddle.  However, if the 
saddle surface is rotated  {\it with an appropriate frequency}   
about the axis normal to 
the surface at the inflection point, 
the particle will be stably confined.
The particle is oscillatory about the origin in both the $x$ and $y$ 
directions.  But it's  oscillation in the $z$ direction 
is restricted to be bounded from below by  some $z_0>0$.

The potential energy can be parametrized as \cite{dawson}
\be
V(x,y,z,t) =  V_x(x,t) + V_y(y,t) +V_z(z,t)  \label{vxyz},
\ee
where
\bea
V_x(x,t) &=&  + \frac{e}{2r_0^2}~{\cal V}(t)~x^2
        \equiv g(t)~ x^2,  \label{vsubx} \\
V_y(y,t) &=&  + \frac{e}{2r_0^2}~{\cal V}(t)~y^2
    \equiv g(t)~ y^2, \\
V_z(z,t) &=&  - \frac{e}{r_0^2}~{\cal V}(t)~z^2
    \equiv g_3(t)~ z^2.  \label{vsubz}
\eea

These potentials can be used to set up the classical motion problem
\bea
{\ddot{x}}_{cl} =  F_x(x,t) & = & -\frac{dV_x(x,t)}{dx}= -2g(t)x_{cl}, 
 \label{xcleq} \\
{\ddot{y}}_{cl} =  F_y(y,t) & = & -\frac{dV_y(y,t)}{dy}= -2g(t)y_{cl},  
    \label{ycleq} \\
{\ddot{z}}_{cl} =  F_z(z,t) & = & -\frac{dV_z(z,t)}{dz}= -2g_3(t)z_{cl}.
    \label{zcleq}
\eea
The solutions to these equations are   
oscillatory Mathieu functions for the bound case \cite{dawson}.  
The oscillatory motion is similar in the $(x,y)$ directions  
but is different in the $z$ direction, since $g(t)\ne g_3(t)$.


\section{The quantum-mechanical Paul trap}

In the quantum mechanical treatment of the Paul trap, one must solve 
the Schr\"odinger equation, which in Cartesian coordinates has the form 
($\hbar=m=1$)
\begin{equation}
{\cal S}\Psi(x,y,z,t)=
\left\{\partial_{xx}+\partial_{yy}+\partial_{zz}+2i\partial_t
-2g(t)(x^2+y^2)-2g_3(t)z^2\right\}\Psi(x,y,z,t)=0.
\label{se1}
\end{equation}
The time-dependent functions $g$ and $g_3$ in Eq. (\ref{se1}) are 
\be
g(t)=  +\frac{e}{2r_0^2}~{\cal V}(t),~~~~
g_3(t)=- \frac{e}{r_0^2}~{\cal V}(t).\label{g2t}
\ee 
where
\be
{\cal V}(t) = {\cal V}_{dc}   -  {\cal V}_{ac}\cos{\omega (t-t_0)}
\label{Epot}
\ee
is the ``dc'' plus ``ac time-dependent'' electric potential that is 
applied between the ring and the end caps.  

Exact solutions for the 1-dimensional, quantum case were first investigated 
in detail by   
Combescure \cite{mc1}.  In general, work has concentrated on the $z$
coordinate, but not entirely \cite{chi}.  Elsewhere \cite{ntnjp}, 
stimulated by the work of Ref. \cite{schrade}, we 
discussed the classical/quantum theory of the Paul trap in the 
z-direction.

The form of the Schr\"odinger equation (\ref{se1}) suggests that, 
in addition to the above Cartesian coordinate system, there is another
natural coordinate system for this problem: the cylindrical coordinate 
system.  Introducing such a change of variables,
\begin{equation}
x=r\cos{\theta},~~~~y=r\sin{\theta},~~~~z=z, \label{cylcord}
\end{equation}
the Schr\"odinger equation becomes
\bea
{\cal S}_{cyl}\Phi(r,\theta,z,t)&=&
\left\{\partial_{rr}+\frac{1}{r}\partial_r
     +\frac{1}{r^2}\partial_{\theta\theta}
+\partial_{zz}+2i\partial_t-2g(t)r^2-2g_3(t)z^2\right\}\Phi(r,\theta,z,t)
       \nonumber \\
   &=& 0.
\label{secyl1}
\eea

In addition, the forms of Eqs. (\ref{se1}) and (\ref{secyl1}) suggest that
we can factorize the solutions and equations into the forms
\be
\Psi(x,y,z,t)=X(x,t)Y(y,t)Z(z,t), \label{ansatz1}
\ee
\be
{\cal S}_xX(x,t)=
\left\{\partial_{xx}+2i\partial_t-2g(t)x^2\right\}X(x,t)=0,
\label{sepxeqn}
\ee
\be
{\cal S}_yY(y,t)=
\left\{\partial_{yy}+2i\partial_t-2g(t)y^2\right\}Y(y,t)=0,
\label{sepyeqn}
\ee
\be
{\cal S}_zZ(z,t)=
\left\{\partial_{zz}+2i\partial_t-2g_3(t)z^2\right\}Z(z,t)=0,
\label{sepzeqn}
\ee
and 
\be
\Phi(x,y,z,t) = \Omega(r,\theta,t)Z(z,t),    \label{ansatz2}
\ee
\be
{\cal S}_{r\theta}\Omega(r,\theta,t)
=\left\{\partial_{rr}+\frac{1}{r}\partial_r
+\frac{1}{r^2}\partial_{\theta\theta}-2g(t)r^2\right\}
\Omega(r,\theta,t)=0.     \label{seprtheqn}
\ee

Being physicists, we tend to blindly go ahead and 
accept this separation, ignoring the
mathematical subtleties in separating coordinates in time-dependent partial
differential equations.  Fortunately, Rod insists on keeping me honest.  
But this procedure does 
turn out to be justified \cite{ntpaulsym}.  So, now we can just go
blindly on.   


\section{Lie symmetries and separable coordinates}

Lie symmetries for the Schr\"odinger equation (\ref{se1}) can be obtained 
by solving the operator equation \cite{wm1,kmnt}  
\begin{equation}
[{\cal S},L]=\lambda(x,y,z,t){\cal S}.\label{lsdf}
\ee
The operator ${\cal S}$ is one of the the Schr\"odinger operators 
we have discussed, 
$L$ is a generator of Lie symmetries, and $\lambda$ is a function of the 
coordinates $x$, $y$, $z$, and $t$.  An operator $L$ has the general form 
\begin{equation}
L=C_0\partial_t+C_1\partial_x + C_2\partial_y +C_3\partial_z+C,
\label{liesym}
\end{equation}
where the coefficient in each term is a function of the coordinates 
and time.  

First we define what are going to be useful separable coordinates:  
\begin{equation}
\x=\frac{x}{\phi^{1/2}(t)},~~~\y=\frac{y}{\phi^{1/2}(t)},~~~
\z=\frac{z}{\phi_3^{1/2}(t)},~~~\tn = t,
\label{Rcarsv}
\end{equation}
and 
\begin{equation}
\rho=\frac{\sqrt{x^2+y^2}}{\phi^{1/2}(t)},~~~
\theta=\sin^{-1}\left(\frac{y}{\sqrt{x^2 + y^2}}\right),~~~
\z=\frac{z}{\phi_3^{1/2}(t)},~~~\tn = t.
\label{newpolcrd1}
\end{equation}
The $t$-dependent functions $\phi(t)$ and $\phi_3(t)$ are  
given by 
\begin{equation}
\phi=2\xi\bar{\xi}~~~~~\phi_3=2\xi_3\bar{\xi_3} , 
\label{crid1}
\end{equation}
where $\{\xi(t),~\bar{\xi}(t)\}$ and   $\{\xi_3(t),~\bar{\xi}_3(t)\}$ are
the complex solutions of the second-order, linear, differential 
equations in time
\begin{equation}
\ddot{\gamma}+2g(t)\gamma=0, ~~~~~~~~
\ddot{\gamma}_3+2g_3(t)\gamma_3=0,\label{gam}
\end{equation}
respectively, and satisfy the Wronskians 
\begin{equation}
W(\xi,\bar{\xi})=W(\xi_3,\bar{\xi}_3)=-i.\label{cw1}
\end{equation}

Equations (\ref{gam}) are the 
same as the classical equations
of motion (\ref{xcleq})-(\ref{zcleq})  for the Paul trap.   
The solutions are the same Mathieu functions.  
For certain values of the parameters 
in the potential (\ref{vxyz}), the solutions are bound, meaning the charged
particle is indeed ``trapped'' \cite{dawson}.  This 
shows a  connection between classical and quantum dynamics. 


\section{Cartesian symmetries}

For this exercise we can concentrate only on the $z$ coordinate, 
since formally the results are
the same for the $x$ and $y$ solutions, with the exception that the 
$\xi_3$'s
and $g_3$'s, etc., lose the subscripts $3$.  [Elsewhere, we will discuss 
the symmetries of the Paul trap in much greater detail \cite{ntpaulsym}.] 

One can find, or simply verify, that the Schr\"odinger equation has the
symmetry operators 
\bea
 &J_{z-}=\xi_3\partial_z-i\dot{\xi}_3z
= +\frac{1}{\sqrt{2}}\left(\frac{\bar{\xi}_3}{\xi_3}
\right)^{\frac{1}{2}}\left[\partial_{\z}+\z\left(1-\lfrac{i}{2}
\dot{\phi}_3\right)\right],& 
       \label{Jz-}  \\
&J_{z+}=-\bar{\xi}_3
\partial_z+i\dot{\bar{\xi}}_3z
= -\frac{1}{\sqrt{2}}\left(\frac{\bar{\xi}_3}{\xi_3}
\right)^{\frac{1}{2}}\left[\partial_{\z}-\z\left(1+\lfrac{i}{2}
\dot{\phi}_3\right)\right].&  
       \label{Jz+} 
\eea
These operators satisfy the nonzero commutation relation 
\begin{equation}
[J_{z-},J_{z+}]=I,~~~ \label{JJcom1} 
\end{equation}
and so form a complex Heisenberg Weyl algebra  $w^c$.  
This  means 
that the operators generate a set of ``number states'' given by 
\bea
J_{z-}Z_{n_z}(z,t) &=& \sqrt{n_z}Z_{n_z-1}(z,t),
        \label{j-Z}\\
J_{z+}Z_{n_z}(z,t) &=& \sqrt{n_z+1}Z_{n_z+1}. \label{j+Z}
\label{zalgebra}
\eea
There is also an $su(1,1)$ algebra, which we will not go into here. 
This is a generalization of the ``squeeze algebra.''

A word of caution.  The states
$Z_{n_z}(z,t)$  
are not to be construed as energy eigenstates.    
$Z_{n_z}$ is a solution to the time-dependent Schr\"odinger 
equation (\ref{sepzeqn}).  
It is generally not an eigenfunction of the 
Hamiltonian.  That is, 
\be
i\partial_tZ_{n_z}= HZ_{n_z}
=i\left[\frac{ i\partial_t{Z}_{n_z}}{Z_{n_z}}\right]Z_{n_z}
\ne [\mathrm{Const}]Z_{n_z}.
\ee

So, to keep John Klauder from getting mad at me, we
use the terminology ``extremal state'' for $Z_{0}$ 
and ``higher-order'' 
states for $Z_{n_z}$.  We restrict the terms ``ground state'' 
and  ``excited state'' for 
problems which allow eigenstates of the Hamiltonian.

Eq. (\ref{j-Z}) is a simple first order differential equation for $Z_0$:
\be
J_{z-}Z_{0}(z,t)  = 0.
\ee
Solving it yields 
\be
Z_0(z,t) = f(\tn) \exp\left\{-\frac{\z^2}{2}
\left[1 - i\frac{\dot{\phi}_{3}}{2}\right]\right\}, \label{z0simp}
\ee
where the proportionality constant, $f$  must be a function of $\tn$.
      
A first idea  might be to 
take this proportionality constant  as the ``normalization constant,'' 
$[\pi\phi_3(t)]^{-1/4}$.  This would conserve the probability, as one
would want for the time-development of a unitary Hamiltonian. 
But this turns out to be incorrect.  Such a
solution {\it does not} satisfy the Schr\"odinger equation 
(\ref{sepzeqn}).  Indeed, putting Eq.     (\ref{z0simp})   into 
Eq.   (\ref{sepzeqn}) yields a first order differential equation in $t$ 
for $f(t)$. The resulting normalized  extremal-state solution is 
\be
Z_{0}(z,t)  =  \left(\pi\phi_{3}\right)^{-1/4}
\left(\frac{\bar{\xi}_3}{\xi_3}\right)^{\frac{1}{4}}   
\exp\left\{-\frac{\z^2}{2}\left[1 - i\frac{\dot{\phi}_{3}}{2}\right]
\right\}.
\label{z0final}
\ee

So why is the phase factor 
$\left({\bar{\xi}_3}/{\xi_3}\right)^{{1}/{4}}$ there?  It is there 
because, in this {\it time-dependent} Schr\"odinger equation, it is the 
necessary generalization of the simple-harmonic oscillator 
ground-state energy exponential,   $\exp[-i\hbar\omega/2]$. (This follows  
since   $\xi_3(t) \rightarrow (2\omega)^{-1/2} \exp[i\omega t]$.)
This phase factor {\it is necessary}  \cite{phase,drt2,nt1}
for Eq. (\ref{z0final}) to be a solution of Eq. (\ref{sepzeqn}).  

Now repeated application of Eq. (\ref{j+Z}) gives 
\be
Z_{n_z}(z,t)= [n_z!]^{-1/2}\left[J_{z+}\right]^{n_z}Z_{0}(z,t).
\label{Z0toZn}
\ee
But the right hand side of Eq. (\ref{Z0toZn}) is 
proportional to \cite{ntpaulsym}
the  Rodrigues formula for the Hermite polynomials 
\cite{mos}.  Using this yields the result 
\bea
Z_{n_z}(z,t) & = & \left(2^{n_z}n_z!\right)^{-1/2}
\left(\pi\phi_{3}\right)^{-1/4}
\left(\frac{\bar{\xi}_3}{\xi_3}\right)
^{\frac{1}{2}\left(n_z+\frac{1}{2}\right)}    \nonumber\\
 &  & ~~~H_{n_z}(\z)~
\exp\left\{-\frac{\z^2}{2}\left[1 - i\frac{\dot{\phi}_{3}}{2}\right]
\right\}.
\label{psiZ}
\eea    

The forms of $X_{n_x}(x,t)$ and $Y_{n_y}(y,t)$ follow immediately by just
changing notation, 
\bea
X_{n_x}(x,t) &=& Z_{n_z \rightarrow n_x}
(z \rightarrow x,\xi_3\rightarrow\xi,\phi_3 \rightarrow \phi, t), \\
Y_{n_y}(y,t) &=&  Z_{n_z \rightarrow n_y}
(z \rightarrow y,\xi_3\rightarrow\xi,\phi_3 \rightarrow \phi, t),
\eea
so that 
\be
\Psi_{n_x,n_y,n_z}(x,y,z,t)=X_{n_x}(x,t)Y_{n_y}(y,t)Z_{n_z}(z,t).
\ee


\section{Polar symmetries}

In what is intuitively interesting, finding the symmetries for the polar
coordinates amounts to considering complex linear combinations 
of the Cartesian operators.   Take the following 
two pairs of operators which form the basis of two Heisenberg-Weyl 
algebras {\cite{kmnt}}:
\bea
a_- = a_+^\dag  &=& \sqrt{\frac{1}{2}}(J_{x-}+iJ_{y-})=\sqrt{\frac{1}{2}}
\left[\xi(\partial_x+i\partial_y)- i\dot{\xi}(x+iy)\right] \nonumber \\
 & = & \lfrac{1}{2}\left(\frac{\xi}{\bar{\xi}}\right)^{1/2}e^{i\theta}
\left[\partial_{\rho}+\frac{i}{\rho}\partial_{\theta}+\rho\left(
1-\lfrac{i}{2}\dot{\phi}\right)\right],\label{aminus2}\\
c_{-} = c_+^\dag &=& \sqrt{\frac{1}{2}}(J_{x-}-iJ_{y-})=\sqrt{\frac{1}{2}}
\left[\xi(\partial_x-i\partial_y)- i\dot{\xi}(x-iy)\right] \nonumber  \\
& = & \lfrac{1}{2}\left(\frac{\xi}{\bar{\xi}}\right)^{1/2}e^{-i\theta}
\left[\partial_{\rho}-\frac{i}{\rho}\partial_{\theta}+\rho\left(
1-\lfrac{i}{2}\dot{\phi}\right)\right].\label{cminus2}
\eea

If we add to the above four operators $I$, the operator 
\be
{\cal K}  =  J_{x+}J_{x-}+J_{y+}J_{y-}+1 
    = a_+a_- + c_+c_- + 1,
\ee
and the angular momentum operator 
\begin{equation}
{\cal L}_z=i(y\partial_x-x\partial_y)=-i\partial_\theta,\label{amop}
\end{equation}
we have that the set  $\{I,a_{\pm},c_{\pm},{\cal K},{\cal L}_z \}$ 
forms a closed algebra.  Further, if we make the transformations 
\bea
f &\equiv & \lfrac{1}{2}\left({\cal K}-{\cal L}_z\right)
= a_+a_-+\lfrac{1}{2},                  \label{fop} \\
d & \equiv & \lfrac{1}{2}\left({\cal K}+{\cal L}_z\right)
= c_+c_-+\lfrac{1}{2}, \label{dop}
\eea
we see that the   
the symmetry algebra has two oscillator subalgebras, 
$\{f,a_{\pm},I\}$ and $\{d,c_{\pm},I\}$, which have only the identity 
operator, $I$, in common.  Therefore, the algebra is 
${\cal G}_{x,y}'=os_a(1)+os_c(1)$, with the two Casimir operators 
${\bf C}_1=a_+a_--f=-\lfrac{1}{2}$ and ${\bf C}_2=c_+c_--d=-\lfrac{1}{2}$.

We restrict ourselves to 
the representations of $os(1)$ that are bounded below; namely, 
$\uparrow_{-1/2}+\uparrow_{-1/2}$.   
Let $\Omega_{n,m}$ be a member of the set of common eigenfunctions of $f$ 
and $d$ spanning the representation space.  Then, 
for $(n,m) \in {\bf Z}_0^+$, we have 
\begin{eqnarray}
& f\Omega_{n,m} = \left(n+\lfrac{1}{2}\right)\Omega_{n,m},~~~~~~
   d\Omega_{n,m}=\left(m+\lfrac{1}{2}\right)\Omega_{n,m}, & 
\label{fdeig}\\*[2mm]
& a_-\Omega_{n,m}=\sqrt{n}\Omega_{n-1,m},~~~~~~
   c_-\Omega_{n,m}=\sqrt{m}\Omega_{n,m-1},& 
\label{acminuseig}\\*[2mm]
& a_+\Omega_{n,m}=\sqrt{n+1}\Omega_{n+1,m},~~~~~~
   c_+\Omega_{n,m} = \sqrt{m+1}\Omega_{n,m+1},& 
\label{acpluseig}  \\
 & {\cal L}_z\Omega_{n,m}=(f-d)\Omega_{n,m}=(m-n)\Omega_{n,m}, & 
\label{Leig}\\*[2mm]
 & {\cal K}\Omega_{n,m}=(d+f)\Omega_{n,m}=(n+m+1)\Omega_{n,m}, &
\label{Keig}
\eea
In order that the spectrum of $d$ and $f$ be 
bounded below, we also have 
\begin{equation}
a_-\Omega_{0,0}=0,~~~~~c_-\Omega_{0,0}=0.\label{extst}
\end{equation}

Eqs.  (\ref{amop}) and (\ref{Leig})  
immediately tell us that the (normalized) $\theta$ dependence of 
the $\Omega_{n,m}(r,\theta,t)$ can be given by 
\be
\Omega_{n,m,}(r,\theta,t)={\cal R}_{n,m}(r,t)\Theta_{n,m}(\theta)
        ={\cal R}_{n,m}(r,t)\frac{\exp[i(m-n)\theta]}{\sqrt{2\pi}}.
\ee
Therefore, 
$\Omega_{0,0}(r,\theta,t)$ is a function {\it only} of $r$ or
$\rho$ and $t$.   In this $(n,m)=(0,0)$ case, both the equations
in  (\ref{extst}) have the same first-order differential form.  
So, we can
solve for $\Omega_{0,0}$ 
similarly to as was done for the $Z_{0}$ solution.
The normalized solution to the
Schr\"odinger equation (\ref{seprtheqn}) is found to be    
\be
\Omega_{0,0}(r,\theta,t)  =  \frac{1}{\sqrt{2\pi}}
\left[\frac{2}{\phi}\right]^{1/2}
\left(\frac{\bar{\xi}}{\xi}\right)^{\frac{1}{2}}
\exp\left\{-\frac{\rho^2}{2}
\left[1 - i\frac{\dot{\phi}}{2}\right]\right\}=X_0(\x,\tn)Y_0(\y,\tn).
\ee

Repeated application of Eqs. (\ref{acpluseig})  gives us the
general result:  
\be 
\Omega_{n,m}(\rho,\theta,t)=(n!m!)^{-1/2}
a_+^nc_+^m\Omega_{0,0}(\rho,\theta,t).  \label{a+c+}
\ee
This turns out also to be a Rodrigues-type formula, although more
complicated than before \cite{ntpaulsym}.   This time it is 
for the generalized Laguerre polynomials \cite{mos}.  
The end result turns out to be 
\bea
\Omega_{n,m}(r,\theta,t) & = & 
\frac{ \exp[i(m-n)\theta]}{\sqrt{2\pi}}
\frac{(-)^kk!}{\left(n!m!\right)^{1/2}}
\left(\frac{2}{\phi(t)}\right)^{1/2}
\left(\frac{\bar{\xi}}{\xi}\right)^{\lfrac{1}{2}(n+m+1)}
\nonumber  \\
 &   &  ~~~~~ \rho^{|n-m|}L_k^{(|n-m|)}(\rho^2)
\exp\left\{-\frac{\rho^2}{2}
\left[1 - i\frac{\dot{\phi}}{2}\right]\right\},    \label{cylwfmn} \\
k & \equiv & \lfrac{1}{2}(n+m-|m-n|).  \label{definek}
\eea

Now let us change to the more standard cylindrical quantum numbers:  
\be
\ell = |\ell_z|, ~~~~~~~~\ell_z = m - n, ~~~~~~~n_r = m+n \ge0.
\label{nrlz}
\ee
We find that 
\bea
\Omega_{n,m}(r,t)&\rightarrow& R_{n_r,\ell}(r,t)
             \Theta_{\ell_z}(\theta), \\
 \Theta_{\ell_z}(\theta)&=& \frac{\exp[i~\ell_z~\theta]}{\sqrt{2\pi}}, \\
 R_{n_r,\ell}(r,t)& = & (-1)^{(n_r-\ell)/2}
\left[\frac{2\left[\frac{n_r-\ell}{2}\right]!}
{\phi(t)\left[\frac{n_r+\ell}{2}\right]!}\right]^{1/2} 
\left(\frac{\bar{\xi}}{\xi}\right)^
{\frac{1}{2}\left(n+m+1\right)}
     \nonumber  \\
 &   & ~~~~~ \rho^{\ell}
L_{\lfrac{1}{2}(n_r-\ell)}^{(\ell)}(\rho^2)
\exp\left\{-\frac{\rho^2}{2}
\left[1 - i\frac{\dot{\phi}}{2}\right]\right\}.  \label{cylwfnl}
\eea
Except for our minus-sign phase convention,
Eq. (\ref{cylwfnl}) resembles the
standard result for the ordinary two-dimensional 
harmonic oscillator \cite{knt}.

This standard solution has the known property that $n_r$ and
$\ell_z$ cannot differ by an odd integer to allow a normalizable 
solution.  [$\lfrac{1}{2}(n_r-\ell)$ must be an integer.] If one
solves the Schr\"odinger equation directly, this falls out, but the
physical reason is not transparent.  However, from the symmetry point of
view, the reason is clear.  The $n$ and $m$ quantum numbers, which reflect
the fundamental symmetries, can be all
non-negative integers.  The $n_r$ and $\ell_z$ quantum numbers reflect a
rotation of the axes by $45^o$.  (See figure \ref{fig:eberlyfig}.)  
So, reaching the allowed 
positions of quantum numbers along these diagonals, scaled by $1/\sqrt{2}$, 
means all integer values of $(n_r,\ell_z)$ are not allowed.    


\begin{figure}[ht]
 \begin{center}
\noindent    
\psfig{figure=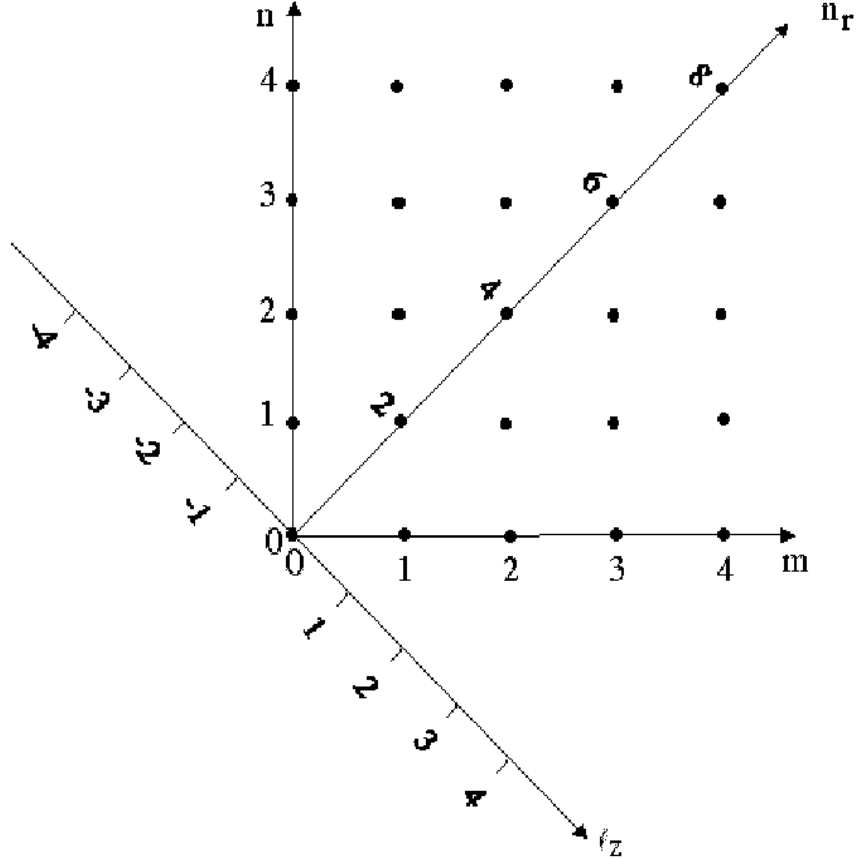,width=4.5in,height=4.5in}
  \caption{A plot of the allowed quantum numbers for the polar
  coordinates of the Paul trap.  Shown are the 
$(n,m)$ quantum numbers of Eqs. (\ref{a+c+}) and (\ref{cylwfmn})  
as well as the 
$(n_r,\ell_z)$ quantum numbers of Eqs. (\ref{nrlz}) and (\ref{cylwfnl}).  
 \label{fig:eberlyfig}}
 \end{center}
\end{figure} 


This is similar in spirit to Pauli's being able to show, by symmetry
methods, that there is an extra conserved quantity, the Runge-Lenz vector, 
om the hydrogen atom.   
This allowed an understanding of the extra degenerate quantum number. 

Finally, we can write out the complete solution in spherical coordinates as 
\be
\Phi_{n_r,\ell_z,n_z}(x,y,z,t)= R_{n_r,\ell}(r,t) \Theta_{\ell_z}(\theta)
             Z_{n_z}(z,t).
\ee


\section*{Appendix:  Getting to know (about) Joe}

Peter Carruthers, who was my advisor at Cornell, had a 
great ability to find talent in physicists.  He also was very perceptive
about the qualities of people, being a well-known lover of life.  For
reasons that will become clear below, let me
give you two examples of Pete's first ability.  

i) We graduate students were amused by
this new, young, hot-shot, assistant professor who was going to publish a
famous paper. We were more amused when he made associate professor.  Later,
I heard of the
move to make him full professor. (From  lack of being there some of these
details must be off.)  It was boiling down to Hans Bethe's call. Pete
figuratively pounded Hans' desk top telling Bethe he had to promote 
this guy. In the end the 
youngster got his full professorship, published his paper (on the
renormalization group), and the rest is the history of Ken Wilson. 

ii) When Pete became division leader at Los Alamos, Pete decided to hire as a
staff member a young guy who had had an unproductive postdoc career -- solely
on the  basis that Pete sensed creativity.  Pete had to hide him in the
Division Office, not a group, because funding could not be justified.  He
just sat around the corner, playing with his HP calculator, getting funny
numbers again and again, leading to the period-doubling, chaos revolution
of Mitchell Feigenbaum.  

I mention this all in the context of when I first heard of Joe Eberly.  
(It was not when I was at Cornell, because then my main contact with
Rochester was with the wild man of the time, Robert Marshak.) It was when
Pete came out to Los Alamos. I asked him if he had seen the new paper by
Eberly and Singh \cite{esingh} on the time-energy uncertainty relation that
had referred to our work.  Pete said he had, and had liked it.  Then he
remarked 
that Eberly ``is very smart.'' Being properly impressed, I figuratively
wrote down Joe's name in my black book of smart guys. Then Pete said 
he's also a nice a guy.  So, Joe also went into the ``nice guy'' book.  
Most of
us, if we are honest, must admit to really wanting to be in the first
book. But Joe is in both books, 
and that is why we are all here today to honor him.


\section*{Acknowledgements}

MMN acknowledges the support of the United States Department of 
Energy.  DRT acknowledges
a grant from the Natural Sciences and Engineering Research Council 
of Canada.



\begin{thebibliography}{}
\bibitem{phase}  The phase factor can also be obtained  {\cite{drt2,nt1}},
by solving the eigenvalue equation
$M_3Z_{n_z}=\left(n_z+\lfrac{1}{2}\right)Z_{n_z}$, where
$M_3=i\left\{\phi_3\partial_t+\lfrac{1}{2}\left(z\partial_z+\lfrac{1}{2}
\right)-\lfrac{i}{4}\ddot{\phi}_3z^2\right\}.$
Then, solving the equation $J_{z-}Z_{0}=0$ will yield the extremal 
state function up to a factor of $(\pi)^{-1/4}$.
\end{thebibliography}
\end{document}